\RequirePackage{fix-cm}
\documentclass[smallextended]{svjour3}       
\smartqed  
\usepackage{graphicx}
\usepackage{amsmath}
\begin{document}

\title{Accurate calculation of the geometric measure of entanglement for multipartite quantum states}

\author{Peiyuan Teng}


\institute{Peiyuan Teng \at
              		Department of Physics\\
              		The Ohio State University\\
              		Columbus, Ohio,43210,USA\\ \email{teng.73@osu.edu}          
}


\maketitle

\begin{abstract}
This article proposes an efficient way of calculating the geometric measure of entanglement using tensor decomposition methods. The connection between these two concepts is explored using the tensor representation of the wavefunction. Numerical examples are benchmarked and compared. Furthermore, we search for highly entangled qubit states to show the applicability of this method.
\end{abstract}

\section{Introduction}
Quantum entanglement is an essential concept in quantum physics and quantum information. Various measures of quantum entanglement have been proposed to characterize quantum entanglement, such as the Von Neumann entanglement entropy. The geometric measure of entanglement\cite{wei2003geometric} has recently gained popularity, owing to its clear geometric meaning. The geometric measure of entanglement was first proposed by Shimony\cite{shimony1995degree}, then generalized to the multipartite system by Barnum and Linden\cite{barnum2001monotones}, and finally examined by Wei and Goldbart, who gave a rigorous proof that it provides a reliable measure of entanglement\cite{wei2003geometric}.

A large amount of research regarding the properties of geometric entanglement has been performed. For example, properties of symmetric states were discussed using the Majorana representation of
such states\cite{aulbach2010maximally}. The geometric measure of entanglement has been discussed theoretically, although few practical numerical evaluation methods are available owing to the complicated structure of a quantum state, whose amplitude is a complex-valued function. A simple way to determine geometric entanglement was given in Ref\cite{PhysRevA.84.022323}, where their method was tested for three or four qubits states with non-negative coefficients.  A problem with this method is that although the overlap will converge, it may not converge to the minimal overlap. Recently, a method to calculate of the geometric measure of entanglement for non-negative tensors was proposed by\cite{hu2016computing}. Our article illustrates a way to numerically calculate the geometric measure of entanglement for any arbitrary quantum state with complex amplitude, which extends the scope of previous numerical methods.

Tensor network theory is currently widely used as a way of simulating physical systems. The idea of tensor network theory is to represent the wavefunction in terms of a multi-indexed tensor, such as the matrix product states (MPS)\cite{verstraete2008matrix}. Therefore, it is also natural to consider the entanglement within this context. Tensor theory  was applied to study quantum entanglement in Ref.\cite{nispherical}\cite{curtef2007conjugate}. Using tensor eigenvalues to study geometric entanglement was discussed in Ref.\cite{ni2014geometric}. The possibility of using tensor decomposition methods to study quantum entanglement was pointed out in Ref.\cite{enriquez2015minimal} in the context of  Minimal Renyi-Ingarden-Urbanik entropy, of which the geometric entanglement is a special case. The asymptotic behavior of the GME for qutrit systems was studied using the PARAFAC tensor decomposition in  Ref.\cite{enriquez2015minimal}.  In this work, we comprehensively study the possibility of using tensor decomposition methods to calculate the geometric measure of entanglement for arbitrary quantum states. Tensor decomposition methods are currently being developed rapidly. By using the new results in tensor decomposition theory, we can not only use the most efficient way to calculate the geometric measure of entanglement, but also gain a deeper understanding of the structure of quantum states from the perspective of theoretical tensor decomposition theory.

To furthermore demonstrate the efficiency of tensor decomposition methods, we conduct a numerical research for maximally and highly entangled quantum states.
Deep understanding of highly entangled multiqubit states is important for quantum information processing. Highly entangled states, such as the cluster states, could be crucial to quantum computers\cite{PhysRevLett.86.5188}.
Highly entangled states are also key parts of quantum error correction and quantum communication\cite{doi:10.1063/1.3511477}. Therefore, searching for highly entangled quantum states is necessary for the development of quantum information science.

In this article, we first review the concept of geometric measure of entanglement and tensor rank decomposition. Then we point out that the spectrum value for a rank-one decomposition is identical to the overlap of wavefunctions.  Our method is capable of calculating an arbitrary (real or complex, symmetrical or non-symmetrical) pure state wavefunction. We also demonstrated that tensor decomposition method can be used to extract the hierarchical structure of a wavefunction. Perfect agreement is found for the examples that we tested.  At last, we use this method to characterize some quantum states. A maximally entangled state that is similar to the HS states is found. In addition, we permormed a numerical search for highly entangled quantum states from four qubits to seven qubits. We provide new examples of such states that are more entangled than a few of currently known states under geometric entanglement.

This article is organized as follows. In Section \ref{marker2}, we mainly focus on the theoretical aspects of tensor theory and entanglement theory. In Section \ref{marker3}, several known examples are calculated to demonstrate the effectiveness of the tensor rank decomposition method. In Section \ref{marker4}, maximally and highly entangled states are searched and discussed.

\section{\label{marker2} Geometric measure of entanglement and tensor decomposition}
\subsection{Geometric measure of entanglement}
The geometric measure of entanglement for multipartite systems was comprehensively examined by Wei and Goldbart in Ref.\cite{wei2003geometric}. Following their notations, we start from a general n-partite pure state 
\begin{equation}
|\psi\rangle=\sum_{p_1,\dots p_n}\chi_{p_1,p_2\dots p_n}|e^1_{p_1}e^2_{p_2}\dots e^n_{p_n}\rangle.
\end{equation}

Define a distance as
\begin{equation}
d=\min_{|\phi\rangle}\||\psi\rangle-|\phi\rangle\|,
\end{equation}

where $|\phi\rangle$ is the nearest separable state, which can be written as
\begin{equation}
|\phi\rangle=\otimes^n_{l=1}|\phi^{l}\rangle.
\end{equation}
$|\phi^{l}\rangle$ is the normalized wavefunction for each party $l$.  A practical choice of the norm could be the Hilbert\textendash Schmidt norm, or equivalently  the Frobenius norm for a tensor, which equals the squared sum of the modulus of the coefficients.

The geometric entanglement can be written as
\begin{equation}
E(|\psi\rangle)=1-|\langle\psi|\phi\rangle|^2.
\end{equation}

It was proved by Wei and Goldbart in Ref.\cite{wei2003geometric} that this measure of entanglement satisfies the criteria of a good entanglement measure.

We can write a wavefunction in the language of tensor. A general n-partite pure state can be written as

\begin{equation}
|\psi\rangle=\sum_{i,j,\dots k}T_{ij\dots k}|ij\dots k\rangle.
\end{equation}

We use tensor $T$ to describe a quantum state and the Frobenius norm of this tensor $\|T\|=1$. The label $i,j,k$ goes from one to the dimension of the Hilbert space of each party.

A direct product state can be written as
\begin{equation}
|\phi\rangle=a_i|i\rangle\otimes a_j|j\rangle\cdots \otimes a_k|k\rangle.
\end{equation}
$a_i|i\rangle$ is a normalized wavefunction for party $i$, here Einstein summation convention is used.

After writing the wavefunction in the language of tensors, we can use the techniques from tensor decomposition theory to calculate the geometric measure of entanglement.

\subsection{Tensor decomposition}

In general, a tensor decomposition method decomposes a tensor into the direct products of several smaller tensors. Moreover, there are two major ways to decompose a tensor. 

One way is the "Tensor Rank Decomposition" or "Canonical Polyadic Decomposition". For an $n$-way tensor, the Tensor Rank Decomposition can be written as

\begin{equation}
T_{mn\cdots p}=\sum\lambda_r a_{rm}\circ a_{rn}\cdots\circ a_{rp}.
\end{equation}
The minimal value of $r$, that can make this expression exact, is called the rank of this tensor.
The Tensor Rank Decomposition can be physically understood as the decomposition of a multipartite wavefunction into the sum of the direct products of the wavefunction from each part. The dyadic product notation "$\circ$" is used, which means that we treat the product as a single tensor. 

Another way to decompose a tensor is the Tucker Decomposition. In some articles, it is called  "Higher Order Singular Value Decomposition (HOSVD)". It can be written as
\begin{equation}
T_{mnp...z}=\sum\lambda_{\alpha\beta\gamma...\omega} a_{\alpha m}\circ a_{\beta n}\circ a_{\gamma p}\cdots\circ a_{\omega z}.
\end{equation}

The Greek letters $\alpha, \beta, \gamma.,..  \omega$ are arbitrary fixed integers.

These two decomposition methods can be regarded as the tensor generalization of the widely used Singular Value Decomposition (SVD) for a matrix.

\begin{equation}
T_{mn}=\sum_{i=1}^{r}\lambda_i a_{im}\circ a_{in}=U S V^{*}.
\end{equation}

$S$ is the singular value matrix. 

Since matrix $S$ is diagonal, different understandings of this singular matrix can lead to different decomposition methods. A detailed discussion of tensor decomposition methods can be found in Ref.\cite{kolda2009tensor}.

The objective function of a rank\textendash $k$ approximation of a tensor can be written as
\begin{equation}
d=\min{\|T_{mn\cdots p}-\sum_{i=1}^{k}\lambda_i a_{im}\circ a_{in}\cdots\circ a_{ip}\|}.
\end{equation}

While for the Tucker decomposition, we can also fix the index range of $\lambda_{\alpha\beta\gamma...\omega}$ and minimize the norm.

When we restrict our $\lambda$ to be a single scalar for both the Tucker Decomposition and the Tensor Rank Decomposition, these two approximations become the same. In another word, they have the same rank\textendash $1$ decomposition. Therefore, our objective function becomes
\begin{equation}
d=\min{\|T_{mn\cdots p}-\lambda a_{m}\circ a_{n}\cdots\circ a_{p}\|}.
\end{equation}

For general quantum states, these tensors and vectors are defined on the complex field ${C}$.
Notice that this objective function has the same form as in our definition of geometric measure of entanglement with $T_{mn\cdots p}=|\psi\rangle$ and $\lambda a_{m}\circ a_{n}\cdots\circ a_{p}=|\phi\rangle$

From a geometric argument, if $\| T_{mn\cdots p}\|=1$ and $\| a_{m}\circ a_{n}\cdots\circ a_{p}\|=1$, then our claim is
\begin{equation}
\boxed{\lambda=\langle\psi|\phi\rangle.}
\end{equation}

This can be understood intuitively: since $\| T_{mn\cdots p}\|$ is a unit vector in our space, and for a fixed $\| a_{m}\circ a_{n}\cdots\circ a_{p}\|$ with unit length, $\| \lambda a_{m}\circ a_{n}\cdots\circ a_{p}\|$ is a line in our vector space ($m\times n\times \cdots p$ dimensional), when we vary $\lambda$. Therefore, our minimization problem can be geometrically understood as finding the minimal perpendicular distance from all the possible direct product lines in the space. Since both vectors are unit vectors, $\lambda$ must equal the angle between these two vectors. Understanding quantum mechanics in the context of geometry has been pointed out in Ref.\cite{brody2001geometric}. Then our geometric entanglement is

\begin{equation}
E(T)=1-| \lambda |^2,
\end{equation}

which is expressed in the language of a tensor.

Tensor decomposition methods have been existing the scientific computing society for some time, moreover, they have been applied to different fields such as statistics and signal processing etc.\cite{kolda2009tensor}.

\subsection{Numerical algorithm}

There are numerous algorithms that can be used for both the Tensor Rank and Tucker decomposition. The Alternate Least Squares algorithm is one of the most popular approaches. We will not discuss the details of the algorithms here. A complete survey of the algorithms can be found in Ref.\cite{kolda2009tensor} and one of the Alternating Least Squares algorithm for Tucker decomposition was given in Ref. \cite{kapteyn1986approach}.

There are also numerous existing code packages that can be utilized on different coding platforms, such as C++ or MATLAB etc. In this article, we use the MATLAB tensor toolbox 2.6 developed by Sandia National Laboratories\cite{bader2012matlab}. This package is already developed and available online.

We must point out a few important facts about the numerical results. Theoretically, both tensor rank decomposition and Tucker decomposition can be used to perform the calculation. In reality, however, some codes are actually written for the set of real numbers $R$.  We need to work in the domain of complex numbers $C$ in order to be able to represent an arbitrary wave function.  Note that different vector spaces will lead to different optimization results. In addition, the decomposed wavefunction may not be normalized. Apractical choice here would be the Alternate Lease Squares algorithm for the Tucker Decomposition ($tucker\_als$) provided in the toolbox.

The Alternate Lease Squares Tucker algorithm involves the following parameters: (i) The initial tensor, i.e., the The tensor that is used to represent quantum states. (ii) The core of the Tucker decomposition, which can be a tensor with any dimension. (In the case of the best rank one approximation or geometric entanglement, this tensor is just a scalar.) (iii)  An optional initial condition, which is used to initialize the iteration and could be set at random. (iv) Optional iteration control parameters.  

After proper normalization of the initial tensor, the output: the best-fitted core scalar is the maximal overlap and the fitted tensors are the corresponding direct product states. This function implements the well-known Higher Order Orthogonal Iteration (HOOI) algorithm for the Tucker approximation, which behaves better than the previous naive HOSVD algorithm\cite{kolda2009tensor}. The details of this algorithm are non-trivial, and can be found in Ref.\cite{doi:10.1137/S0895479898346995}. The original HOOI paper was formulated in terms of a real tensor, but as pointed out by the authors of Ref.\cite{doi:10.1137/S0895479898346995}, this algorithm equally applies to a complex tensor. Moreover, our numerical study also shows the applicability to quantum states with complex amplitudes.

From the viewpoint of tensor decomposition theory, we can see that previous work about the numerical evaluation of geometric entanglement\cite{PhysRevA.84.022323} is a special case of a naive HOSVD algorithm, which was used at an early stage of Tucker decomposition. The problem with the naive HOSVD in Ref. \cite{PhysRevA.84.022323} is that although the wavefunction overlap  converges, the converged overlap value may not be the minimal overlap in the Hilbert space, see section 4.2 in Ref. \cite{kolda2009tensor} . The HOOI algorithm is designed to minimize the norm and, therefore, it gives the correct result for the geometric measure of entanglement. Another practical point is that the solution may not be unique, and the result may be trapped in  locally minimal state\cite{kolda2009tensor} . Therefore, for consistency, it is better to examine the initial conditions for all the calculations.

\section{\label{marker3} Numerical evaluation of geometric measure of entanglement using alternate least square algorithm}

\subsection{Geometric measure of entanglement for symmetric qubits pure states}

We would like to benchmark the results given by Wei and Goldbart in Ref.\cite{wei2003geometric}. 

Considering a general $n$ qubit symmetric state

\begin{equation}
|S(n,k)\rangle=\sqrt{\frac{k!(n-k)!}{n!}}\sum_{permutations}|0\cdots 01\cdots 1\rangle.
\end{equation}

$k$ is the number of $|0\rangle$s, and $n-k$ is the number of $|1\rangle$s.

The overlap is given by
\begin{equation}
\Lambda=\sqrt{\frac{n!}{k!(n-k)!}}(\frac{k}{n})^{\frac{k}{2}}(\frac{n-k}{n})^{\frac{n-k}{2}}.
\end{equation}

In Table 1, we use $\Lambda$ to denote the theoretical results and the $\lambda$ to label the numerical ones.
\begin{table}
	\caption {\textbf{Overlaps for n-partite qubit systems}}
	\begin{center}
		
		\begin{tabular}{| c | c | c | c |}
			\hline
			n value & k value & $\Lambda$ theoretical & $\lambda$ numerical \\ \hline
			4 & 0 & 1 & 1.0000  \\ \hline
			& 1 & 0.6495 & 0.6495  \\ \hline
			& 2 & 0.6124 & 0.6124 \\  \hline
			& 3 & 0.6495 & 0.6495   \\ \hline		
			5 & 0 & 1 & 1.0000 \\ \hline
			& 1 & 0.6400 & 0.6400    \\ \hline
			& 2 & 0.5879 & 0.5879   \\  \hline
			& 3 & 0.5879 & 0.5879   \\ \hline		  
			& 4 & 0.6400 & 0.6400  \\ \hline  
			6 & 0 & 1 & 1.0000 \\ \hline
			& 1 & 0.6339 & 0.6339   \\ \hline
			& 2 & 0.5738 & 0.5738 \\  \hline
			& 3 & 0.5590 & 0.5590  \\ \hline		  
			& 4 & 0.5738 & 0.5738 \\ \hline	
			& 5 & 0.6339 & 0.6339  \\ \hline
		\end{tabular}
		\bigskip
		
		Comparison between theoretical value $\Lambda$ and the calculation using tensor decomposition $\lambda$. Alternate Lease Square method for Tucker decomposition is used. 
	\end{center}
\end{table}

We test the overlaps for both methods up to 6-partite systems, i.e. 6-way tensors. Agreements are found.

\subsection{Geometric measure of entanglement for combinations of three qubits W states}

Assuming we have a superposition of two W states

\begin{equation}
|\psi\rangle=\sqrt{s}|S(3,2)\rangle+\sqrt{1-s}e^{i\phi}|S(3,1)\rangle\\=
\sqrt{s}|W\rangle+\sqrt{1-s}e^{i\phi}|\widetilde{W}\rangle.
\end{equation}

We can gauge out the factor $\phi$ by changing basis without affecting the entanglement. The geometric measure of entanglement of this state is given by, see Ref.\cite{wei2003geometric}
\begin{equation}
E=1-\Lambda^2.
\end{equation}
With (notice that there is a typo in\cite{wei2003geometric} for this equation)
\begin{equation}
\Lambda=\frac{\sqrt{3}}{2}[\sqrt{s}cos\theta(s)+\sqrt{1-s}sin\theta(s)]sin2\theta(s).
\end{equation}

$t=tan\theta$, where $t$ is the real root of the equation

\begin{equation}
\sqrt{1-s}t^3+2\sqrt{s}t^2-\sqrt{1-s}t-\sqrt{s}=0.
\end{equation}

\begin{figure}
	\includegraphics[width=\linewidth]{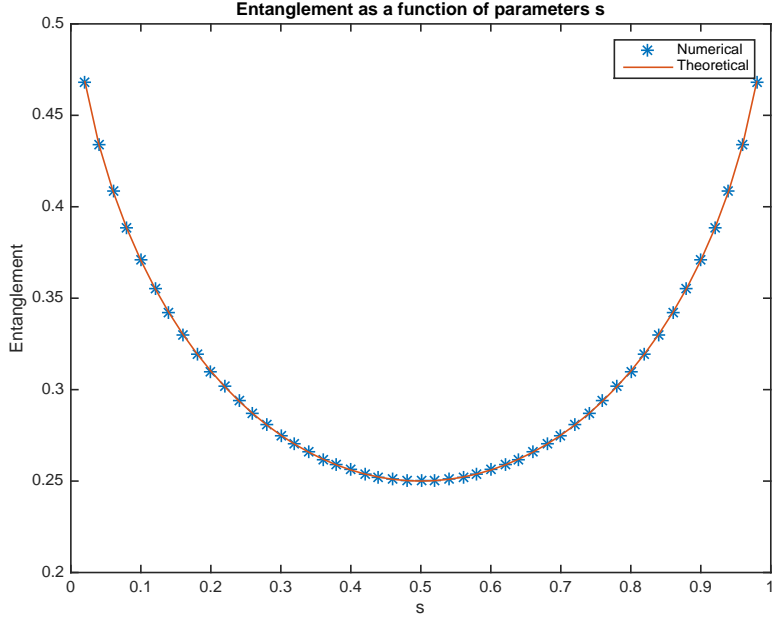}
	\caption{Entanglement as a function of s using tensor decomposition.}
	\label{fig:fig1}
\end{figure}

Perfect agreement is found, see Figure \ref{fig:fig1}.

\begin{figure}
	\includegraphics[width=\linewidth]{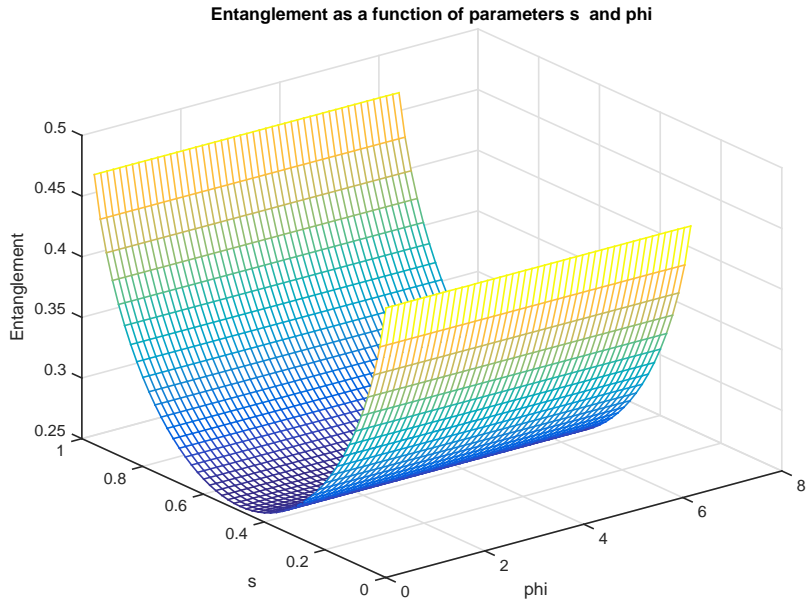}
	\caption{Decomposition for complex tensors. Entanglement for two parameters using tensor decomposition, entanglement is not affected by $\phi$.}
	\label{fig:fig2}
\end{figure}

For a general complex wavefunction
\begin{equation}
|\psi\rangle=|W\rangle+\sqrt{1-s}e^{i\phi}|\widetilde{W}\rangle.
\end{equation}
Tensor decomposition method can indeed capture the complex $\phi$ factor and reflect the fact that this factor does not affect the entanglement. See Figure \ref{fig:fig2}.

\subsection{Geometric measure of entanglement for $d$-level system (qudits)}

Up till now, the index of our tensor has a range of two, which corresponds to a qubit system. We can obviously use a tensor that has a larger index range which corresponds to a $d$-level system.

For example, we have a symmetric state with $n$ parts, for simplicity we assume that one part is in state $d-1$, the other parts are all in state $0$, and our wavefunction is a symmetric sum of all these possible state.
\begin{equation}
|S(n,d)\rangle=\sqrt{\frac{(n-1)!}{n!}}\sum_{permutations}|0\cdots 0(d-1)\rangle.
\end{equation}
The overlap is given by, from Ref.\cite{wei2003geometric}

\begin{equation}
\Lambda=\sqrt{\frac{n!}{(n-1)!}}(\frac{1}{n})^{\frac{1}{2}}(\frac{n-1}{n})^{\frac{n-1}{2}},
\end{equation}

which is independent of $d$.

In Table 2, we use $\Lambda$ to denote the theoretical results and the $\lambda$ to label the numerical ones.

\begin{table}
	\caption {\textbf{Overlaps for n-partite qudit systems}}
	\begin{center}
		
		\begin{tabular}{| c | c | c | c |}
			\hline
			n value & d value & $\Lambda$ theoretical & $\lambda$ numerical \\ \hline
			4 & 4 & 0.6495 &  0.6495  \\ \hline
			& 10 &  & 0.6495  \\ \hline
			& 100 &  & 0.6495 \\  \hline
			& 200 &  & 0.6495   \\ \hline		
			5 & 5 & 0.6400 & 0.6400    \\ \hline
			& 10 &  & 0.6400   \\ \hline
			& 20 &  & 0.6400  \\  \hline
			& 50 &  & 0.6400   \\ \hline	
			6 & 6 & 0.6339
			&  0.6339
			\\ \hline
			& 10 &  & 0.6339
			\\ \hline
			& 20 &  & 0.6339
			\\  \hline

		\end{tabular}
		\bigskip
		
		Comparison between theoretical value $\Lambda$ and the calculations using tensor decomposition $\lambda$ for qudit n-partite states.
	\end{center}
\end{table}

We tested the overlaps of qudit systems up to 6-partite system, i.e. 6-way tensors. Each tensor is tested up to a bond dimension of 50 to 200. The largest tested tensor has a bond  dimension of $50^5$ owing to the restriction of computational power. Agreements are found. Our results show that tensor decomposition method is capable of dealing with qudit systems.

\subsection{Hierarchies of Geometric measure of entanglement}

In general, hierarchies of geometric measure of entanglement refers to the structure of the distances from a quantum state to the K-separable states. For example, for a general pure state, some parts of the system is entangled while the wavefunction can still be written as the direct products of some larger parts. A detailed discussion of the hierarchies can be found in Ref. \cite{blasone2008hierarchies}.

We need to point out that this hierarchy structure of entanglement is quite natural to understand in the context of tensor theory. For a general tensor $T_{ij\cdots k}$, we can combine the first two index together $T_{(ij)\cdots k}$ and write $ij$ as a single index $l$, which means that we treat them as one part. To calculate the entanglement for this partition, what we need is to find the entanglement for the tensor  $T_{l\cdots k}$.  It is easy to see that different partitions are equivalent to different ways of combining tensor indices. Therefore, it is natural to understand the hierarchies in the language of tensor.

For example, if we have a quantum state which has three parties and can be written as a three index tensor $T_{2, 3, 4}$, i.e. the dimension of the Hilbert space  of each party is 2, 3, 4 respectively. Naturally, we can consider a 2-separable state where one party has a Hilbert dimension of 6 and another party has a dimension of 4. Writing in the language of tensors, for  $T_{2, 3, 4}$, we can rewrite the first to labels as one single label which has a bond dimension of $2\times 3=6$, simply by rewriting each 2 by 3 matrix as a 6 dimensional vector. Therefore, by combining two indices we get a new tensor $T_{6, 4}$. Although these two tensors have a one to one map, the tensor decomposition structure has changed, therefore, we can calculate two parties geometric entanglements by different ways of combining indices.  

We would like to calculate the hierarchies for the 5-qubits W state, and compare with the results in Ref. \cite{blasone2008hierarchies}.

\begin{equation}
|W\rangle=
\sqrt{\frac{1}{5}}(|00001\rangle+|00010\rangle+|00100\rangle+|01000\rangle+|10000\rangle).
\end{equation}

\begin{table}
	\caption {\textbf{Hierarchies of 5-qubits W state}}
	\begin{center}
		
		\begin{tabular}{| c | c | c | c | c |}
			\hline
			Partition & Tensor size & $\lambda$ numerical &  E  & E from\cite{blasone2008hierarchies} \\ \hline
			1,4 & $2\times 16$ & 0.8944 & 0.2000 & 0.200   \\ \hline
			2,3 & $4\times 8$ & 0.7745  & 0.4001  &  0.400  \\ \hline
			1,1,3 & $2\times 2\times 8$ & 0.7745 & 0.4001 & 0.400   \\ \hline			
			1,2,2 & $2\times 4\times 4$ &0.6761 & 0.5429 &  0.543  \\ \hline
			1,1,1,2 &$2\times2\times 2\times 4$ &0.6639 & 0.5592 & 0.559  \\ \hline	
			1,1,1,1,1 & $2\times2\times 2\times 2\times 2$ &0.6400 & 0.5904  & 0.590  \\ \hline					
		\end{tabular}
		\bigskip
		
		Comparison of hierarchies using tensor decomposition method.
	\end{center}
\end{table}

We found agreements between these results, see Table 3 for more details. Tensor decomposition method is capable of finding the hierarchical structure of a quantum state.

Tensor decomposition method can also be used to find the Hierarchies of geometric entanglement for non-symmetric states.

For example,

\begin{equation}
|\psi_W^3\rangle=
N_3(\gamma_1|001\rangle+\gamma_2|010\rangle+\gamma_3|100\rangle).
\end{equation}

The theoretical value of overlap square is found to be\cite{blasone2008hierarchies}

\begin{equation}
\Lambda^2(i|j,k)=
N_3^2max[\gamma_i^2,\gamma_j^2+\gamma_k^2].
\end{equation}
Where $i,j,k$ are labels for different parties.

With $\gamma_1=1$,$\gamma_2=2$,$\gamma_3=3$, the overlap square is found to be at $0.6428$ using tensor decomposition method, which is in perfect agreement with the theoretical value.

For another example,

\begin{equation}
|\psi_W^4\rangle=
N_4(\gamma_1|0001\rangle+\gamma_2|0010\rangle+\gamma_3|0100\rangle+\gamma_4|1000\rangle).
\end{equation}

The theoretical value of overlap square is found to be\cite{blasone2008hierarchies}

\begin{equation}
\Lambda^2(i,j|k,l)=
N_4^2max[\gamma_i^2+\gamma_j^2,\gamma_k^2+\gamma_l^2].
\end{equation}

With $\gamma_1=1$,$\gamma_2=2$,$\gamma_3=3$,$\gamma_4=4$, the overlap square is found to be at $0.8333$ using tensor decomposition method, which is also in perfect agreement with the theoretical value.

\section{\label{marker4} Searching for highly entangled states and maximally entangled states.}

Deep understanding of highly entangled multiqubit states is important for quantum information processing. In this section, we discuss several maximally or highly entangled quantum states.
 
 \subsection{Bounds on the geometric measure of entanglement}
 
 By exploiting the correspondence between the geometric measure of entanglement and best rank-one approximation, properties of the geometric measure of entanglement, such as the upper bound, can be acquired.
 
For example, considering a quantum state that can be represented by a real tensor $T$. Assuming the party number is $m$, and the dimension of the each party is given by $2\leq n_1\leq n_2\leq \cdots \leq n_m$. Then the overlap in the real space satisfies

\begin{equation}
\frac{1}{\sqrt{n_1 n_2 \cdots  n_{m-1}}} <\lambda \leq 1.
\end{equation}

Therefore,

\begin{equation}
 0\leq E<1- \frac{1}{n_1 n_2 \cdots  n_{m-1}}.
\end{equation}

Based on the states that we tested,   we could say that this bond is valid. It is not clear whether or when this bound is exact. For mathematical details, please see Ref.\cite{LiqunQi}.

\subsection{Maximally entangled four qubits states}

The four qubits quantum state, Higuchi- Sudbery (HS) state, is conjectured to be maximally entangled\cite{Higuchi2000213}.

We consider a family of Higuchi- Sudbery states $|HS\rangle_t$, where $w=e^{\frac{2\pi i}{3}}$ corresponds to the previously discovered HS state.
\begin{equation}
|HS\rangle_t=
\sqrt{\frac{1}{6}}[|0011\rangle+|1100\rangle+ w(|1010\rangle+|0101\rangle)+w^2(|1001\rangle+|0110\rangle)],
\end{equation}

\begin{figure}
	\includegraphics[width=\linewidth]{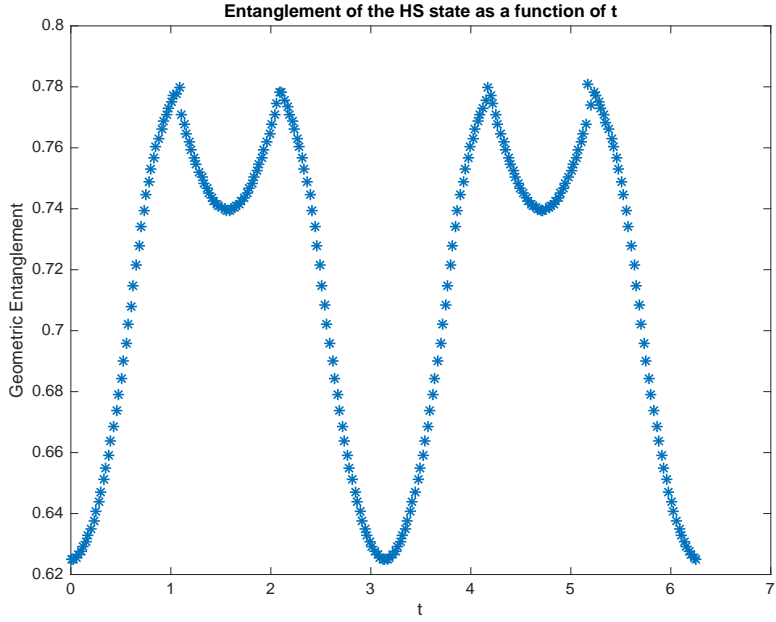}
	\caption{Entanglement of the HS family of states as a function of t, $w=e^{t i}$. The maximal entanglement, $E_{max}=0.7778$.}
	\label{fig:fig3}
\end{figure}

In Figure \ref{fig:fig3}, we show the evolution of the geometric entanglement as a function of $w$. As expected, $E$ has a maximum at $w=e^{\frac{2\pi i}{3}}$. We also notice that the state at $w=e^{\frac{\pi i}{3}}$ has the same entanglement as the $|HS\rangle$ state. Therefore, we have  numerically discovered a few four qubits states that is maximally entangled. However,  we should point out that these states might be equivalent under local unity transformations.

We searched complex four qubits state using Monte Carlo sampling with 100000 samples.  We did not find any four qubits quantum states with a higher geometric entanglement, therefore, the  $|HS\rangle$ is likely to be the four qubits state with the highest entanglement under geometric entanglement measure.

\subsection{Highly entangled four qubits states}

The L state maximizes the average Tsallis $\alpha$-entropy of the partial trace for $\alpha>0$\cite{1742-6596-698-1-012003}. While, surprisingly, we find that this state has a constant geometric entanglement $E=0.6667$ with respect to changing $w$.

\begin{equation}
|L\rangle_t=
\sqrt{\frac{1}{12}}[(1+w)(|0000\rangle+|1111\rangle)+(1-w)(|0011\rangle+|1100\rangle)+w^2(|0101\rangle+|0110\rangle+|1001\rangle+|1010\rangle)].
\end{equation}

The $|BBSB_4\rangle$ is found to be a highly entangled state with respect to a certain measure \cite{BSSB}.
\begin{equation}
|BSSB_4\rangle=
\sqrt{\frac{1}{8}}[|0110\rangle+|1011\rangle+i(|0010\rangle+|1111\rangle)+ (1+i)(|0101\rangle+|1000\rangle)].
\end{equation}

\begin{figure}
	\includegraphics[width=\linewidth]{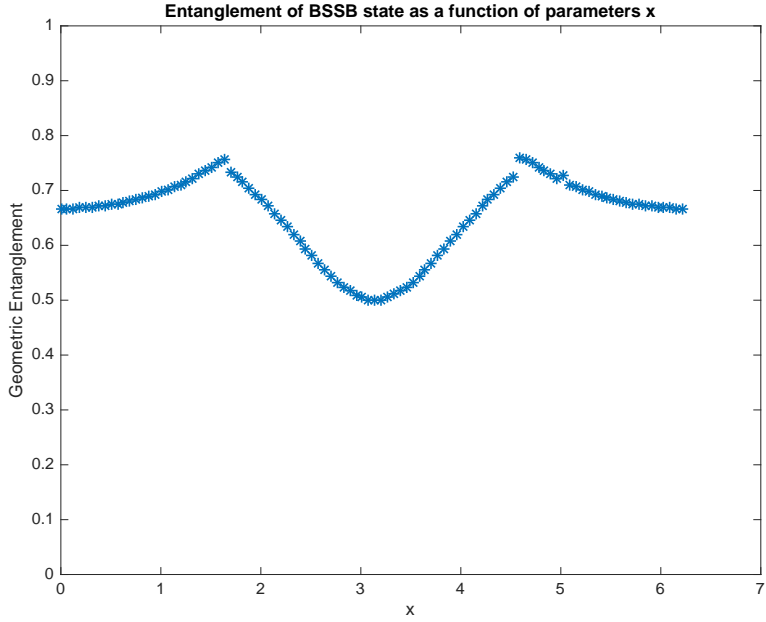}
	\caption{Entanglement of the BSSB family of states as a function of x, $w=e^{x i}$. The entanglement at $w=i$ is $E_{BSSB4}=0.7500$.}
	\label{fig:fig4}
\end{figure}

 Our result shows that it is a local minimum under a family of $|BSSB_4\rangle_t$ states, at $w=i$ with $E_{BSSB4}=0.7500$, see Figure \ref{fig:fig4}. 

\begin{equation}
|BSSB_4\rangle_t=
\sqrt{\frac{1}{8}}[|0110\rangle+|1011\rangle+w(|0010\rangle+|1111\rangle)+ (1+w)(|0101\rangle+|1000\rangle)].
\end{equation}

In addition to the highly entangled state listed above, we provide a list of highly entangled four qubits states, based on our numerical search. The states with real integer coefficients are relatively easy to prepare in experiment. These states have the same entanglement as the $|BBSB_4\rangle$  state.

\begin{equation}
|\phi_{4,1}\rangle=
\frac{1}{2}(|0000\rangle+|1110\rangle+|0101\rangle+|1011\rangle).
\end{equation}

\begin{equation}
|\phi_{4,2}\rangle=
\frac{1}{2}(|1100\rangle+|0010\rangle+|0101\rangle+|1011\rangle).
\end{equation}

\begin{equation}
|\phi_{4,3}\rangle=
\frac{1}{2}(|1000\rangle+|0110\rangle+|0001\rangle+|1111\rangle).
\end{equation}

\begin{equation}
|\phi_{4,4}\rangle=
\frac{1}{2}(|0100\rangle+|0010\rangle+|1001\rangle+|1111\rangle).
\end{equation}

\begin{equation}
|\phi_{4,5}\rangle=
\frac{1}{2}(|0110\rangle+|1010\rangle+|0001\rangle+|1101\rangle).
\end{equation}

\begin{equation}
|\phi_{4,6}\rangle=
\frac{1}{2}(|0010\rangle+|1110\rangle+|0101\rangle+|1001\rangle).
\end{equation}

\begin{equation}
|\phi_{4,7}\rangle=
\frac{1}{2}(|0000\rangle+|1100\rangle+|0011\rangle+|1111\rangle).
\end{equation}
 All the states above have a overlap of $\lambda=0.5$  and a geometric entanglement of $E=0.75$.
 
 All the $|\phi\rangle$ states in this paper are constructed and searched using Monte Carlo Sampling. We start with a several index tensor and random initialize each element of tensor to zero or one. Practically, the number of 1s in each tensor is fixed under each Monte Carlo process, althought different value is used in different run. Then we normalize each tensor and calculate the geometric entanglement. Using a large number of samples, the tensor with the largest geometric entanglement is recorded.
 
 \subsection{Highly entangled five qubits states}
 
 The $|BBSB_5\rangle$ state is found to be a highly entangled five qubits state\cite{BSSB}.
 
 \begin{equation}
|BBSB_5\rangle=
\sqrt{\frac{1}{8}}(|00001\rangle-|00010\rangle+|01000\rangle-|01011\rangle  +|10001\rangle+|10010\rangle+|11100\rangle+|11111\rangle).
\end{equation}
 
The geometric entanglement is 0.7500. Our search find a new state $|\phi_{5,1}\rangle$, which is more entangled than
$|BBSB_5\rangle$ under the measure of geometric entanglement.
 \begin{equation}
|\phi_{5,1}\rangle=
\sqrt{\frac{1}{6}}(|00000\rangle+|01100\rangle+|10010\rangle+|11001\rangle+|00111\rangle+|11111\rangle).
\end{equation}
For $|\phi_{5,1}\rangle$, the overlap is $\lambda=0.4329$   with entanglement $E=0.8126$.
 
\begin{equation}
|\phi_{5,2}\rangle=
\sqrt{\frac{1}{8}}(|11000\rangle+|01100\rangle+|10010\rangle+|10110\rangle  +|00001\rangle+|01001\rangle+|00111\rangle+|11111\rangle).
\end{equation}

For $|\phi_{5,2}\rangle$, the overlap is $\lambda=0.500$  with entanglement $E=0.7500$, which is the same as $|BBSB_5\rangle$.
 
 \subsection{Highly entangled six and seven qubits states}
 
We provide two examples of six qubits state.

\begin{equation}
 |\phi_{6,1}\rangle=
 \sqrt{\frac{1}{7}}(|100000\rangle+|011000\rangle+|011110\rangle+|101110\rangle+|101001\rangle+|110101\rangle+|000011\rangle).
\end{equation}
 
For $|\phi_{6,1}\rangle$, the overlap is $\lambda=0.3780$   with entanglement $E=0.8571$.
 
 \begin{equation}
  |\phi_{6,2}\rangle=
  \sqrt{\frac{1}{8}}(|11000\rangle+|001100\rangle+|010110\rangle+|100110\rangle  +|001001\rangle+|100101\rangle+|111101\rangle+|101011\rangle ).
  \end{equation}
   
For $|\phi_{6,2}\rangle$   the overlap is $\lambda=0.3954$   with entanglement $E=0.8436$.
   
Notice that our six qubits states are more simple than the state found in Ref.\cite{1751-8121-40-44-018}.
 
For seven qubits states, we found
 
\begin{multline}
|\phi_{7,1}\rangle=
\sqrt{\frac{1}{10}}(|0110000\rangle+|0011000\rangle+|1100100\rangle+|0001100\rangle+|1110010\rangle\\+|1001010\rangle+|1101001\rangle+|1010101\rangle+|0000011\rangle+|1111111\rangle).
\end{multline}

For $|\phi_{7,1}\rangle$, the overlap is $\lambda=0.3162$   with entanglement $E=0.9000$.
 
\begin{multline}
|\phi_{7,2}\rangle=
\sqrt{\frac{1}{11}}(|0110000\rangle+|0000100\rangle+|1100100\rangle+|1011100\rangle+|1001010\rangle\\+|0011110\rangle+|0101101\rangle+|1110011\rangle+|0000011\rangle+|0011011\rangle+|1010111\rangle).
\end{multline}

For $|\phi_{7,2}\rangle$, the overlap is $\lambda=0.3183$   with entanglement $E=0.8987$.

Notice the geometric entanglement of all the states in this section is invariant under local unitary transformation of each party. Therefore, we can get other state by applying a rotation on each qubit.
 
\section{Discussions}
 \subsection{Geometric measure of entanglement for many-body systems}
The geometric measure of entanglement defined above is for finite quantum states. For many-body systems, we can define geometric entanglement per site, by using the overlap between an entangled state an a direct product state of every site. For a 1-D system, the ground state can be written as a Matrix Product State (MPS). Assuming translational symmetry, we can efficiently calculate geometric entanglement per site based on the local structure of the MPS representation. Remarkably, the geometric entanglement structure for a translational symmetric many-body system is more simpler for a finite state space. For certain 1-D systems, analytical solutions exist. The details of the process discussed above can be found in Ref. \cite{PhysRevA.81.062313}.

 Recently, research has been performed for 2-D systems. For a 2-D translational invariant quantum many-body system, the ground state can be represented as an infinite Projected Entangled Pair States (iPEPS).  Following the same procedure used for the 1-D case, geometric entanglement per site can be calculated by contracting over the tensor network representation of the overlap coefficient. The overlap is dominated by the largest singular value of the representation tensor treated as a matrix. The largest singular value of a matrix is the same as the overlap coefficient of the best rank-one approximation of the tensor discussed in this paper.  We can use this overlap and geometric entanglement to discover phase transition for a 2-D many-body system (see details in Ref.\cite{PhysRevA.93.062341}). For a 2-D system, an iPEPS tensor can be represented as a matrix and therefore, it is still easy to calculate. Our method is potentially beneficial to the tensor representation of a 3-D or higher dimensional system, although a realistic tensor representation for 3-D quantum systems is beyond the current computational power.

 \subsection{Several comments}
A topic that we have not discussed in this article is the calculation of the geometric measure of entanglement for mixed states. It is known that the entanglement curve for a mixed state is the convex hull for the corresponding pure state. After numerically calculation the entanglement surface of the pure states, it should be straightforward to calculate the convex hull geometrically using numerical methods.

A subtle detail that we should stress is that the tensor decomposition may be trapped in a numerical metastable state if the initial conditions are not properly set. Therefore, for a reliable calculation, great care should be given to the initial conditions to avoid erroneous results.

Tensor decomposition theory is currently still under development and therefore, some theoretical aspects of its properties are still unknown. It will be interesting if new developments of tensor decomposition theory shed some light on quantum theory and quantum information theory.

It will be interesting to explore the restrictions of this method. It is known that the calculation of the best rank-one approximation of a tensor is NP-hard\cite{hillar2013most}, which was also proved in Ref.\cite{huang2014computing}. Therefore, it is difficult to calculate a geometric measure of entanglement for a rather large quantum system. Our method is easy to implement, and is based on existing code packages. Based on existing calculation software such as MATLAB.  Convex hull (Convex envelop) can also be constructed in MATLAB, to represent the entanglement of mixed quantum states, see \cite{wei2003geometric} for details.

\section{Conclusions}

In this article, we established the connection between tensor decomposition theory and the geometric measure of entanglement. We found agreements between theoretical and numerical method. Furthermore, we searched and characterized several quantum states with high entanglement. We illustrated that the tensor decomposition method is an efficient and accurate method for the calculation of the geometric measure of entanglement.

\begin{acknowledgements}
I want to give my sincere thanks to the Ohio State University Physics Department who financially supported my study. This research did not receive any specific grant from funding agencies in the public, commercial, or not-for-profit sectors.

\end{acknowledgements}

\bibliographystyle{spphys} 
\bibliography{GE}

\end{document}